\documentclass[review]{elsarticle}

\usepackage{multirow}
\usepackage[cmintegrals]{newtxmath}
\usepackage{graphicx}
\usepackage{algorithm}
\usepackage{algorithmic}
\usepackage{subfigure}
\usepackage{wrapfig}
\usepackage{multirow}
\usepackage{color}
\usepackage{bbding}

\journal{Journal of \LaTeX\ Templates}

\begin{document}
	
	\begin{frontmatter}

		\title{Feature Pyramid Attention based Residual Neural Network for Environmental Sound Classification}
		
		
		
		
		\author[mymainaddress]{Liguang Zhou}
		
		\author[mysecondaryaddress]{Yuhongze Zhou}
		\author[mymainaddress]{Xiaonan Qi}
		\author[mymainaddress]{Junjie Hu}
		
		\author[mymainaddress]{Tin Lun Lam \corref{mycorrespondingauthor}}
		\cortext[mycorrespondingauthor]{Corresponding author}
		\ead{tllam@cuhk.edu.cn}
		
		\author[mymainaddress]{Yangsheng Xu}
		
		\address[mymainaddress]{Shenzhen Institute of Artificial
			Intelligence and Robotics for Society, The Chinese University of Hong Kong,
			Shenzhen }
		\address[mysecondaryaddress]{Mcgill University }

		\begin{abstract}
			Environmental sound classification (ESC) is a challenging problem due to the unstructured spatial-temporal relations that exist in the sound signals. Recently, many studies have focused on abstracting features from convolutional neural networks while the learning of semantically relevant frames of sound signals has been overlooked. To this end, we present an end-to-end framework, namely feature pyramid attention network (FPAM), focusing on abstracting the semantically relevant features for ESC. We first extract the feature maps of the preprocessed spectrogram of the sound waveform by a backbone network. Then, to build multi-scale hierarchical features of sound spectrograms, we construct a feature pyramid representation of the sound spectrograms by aggregating the feature maps from multi-scale layers, where the temporal frames and spatial locations of semantically relevant frames are localized by FPAM. Specifically, the multiple features are first processed by a dimension alignment module. Afterward, the pyramid spatial attention module (PSA) is attached to localize the important frequency regions spatially with a spatial attention module (SAM). Last, the processed feature maps are refined by a pyramid channel attention (PCA) to localize the important temporal frames. To justify the effectiveness of the proposed FPAM, visualization of attention maps on the spectrograms has been presented. The visualization results show that FPAM can focus more on the semantic relevant regions while neglecting the noises. The effectiveness of the proposed methods is validated on two widely used ESC datasets: the ESC-50 and ESC-10 datasets. The experimental results show that the FPAM yields comparable performance to state-of-the-art methods. A substantial performance increase has been achieved by FPAM compared with the baseline methods.
		\end{abstract}
		
		
	\end{frontmatter}
	
	
	\section{Introduction}
	Environmental sound classification (ESC) is an important research direction for scene understanding. Via ambient sound of the environment, the mobile agent is capable of perceiving the environment that it is traveling by. ESC has been used in various applications, including robot hearing \cite{ren2016sound}, relative bearing estimation of robots \cite{basiri2016board}, smart home automation systems \cite{Chen2013,wang2015robust,do2021soham}, elderly care \cite{do2018rish}, automatic construction \cite{akbal2022learning}, and robot discovery of the auditory scene \cite{martinson2007robotic}.

	
	\begin{figure*}[t]
		\begin{center}
			\subfigure[glass breaking]{\includegraphics[width=3cm]{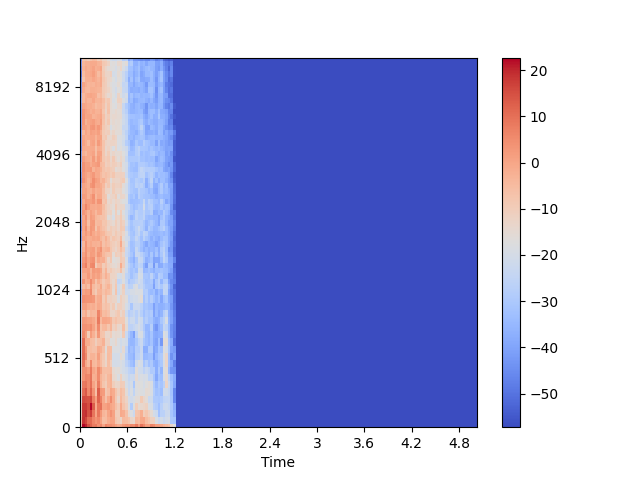}}
			\subfigure[can opening]{\includegraphics[width=3cm]{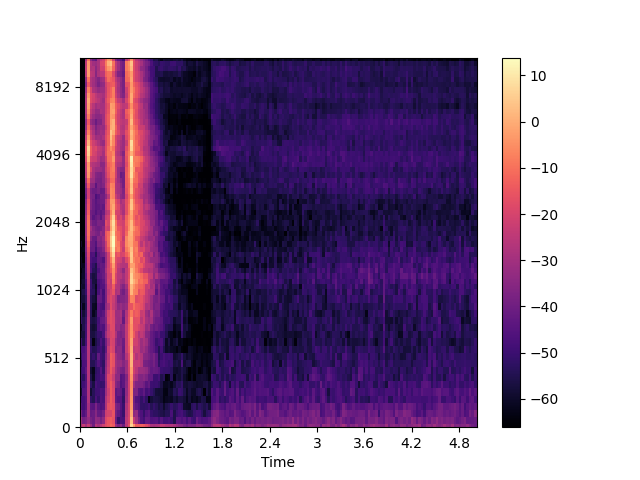}}
			\subfigure[airplane]{\includegraphics[width=3cm]{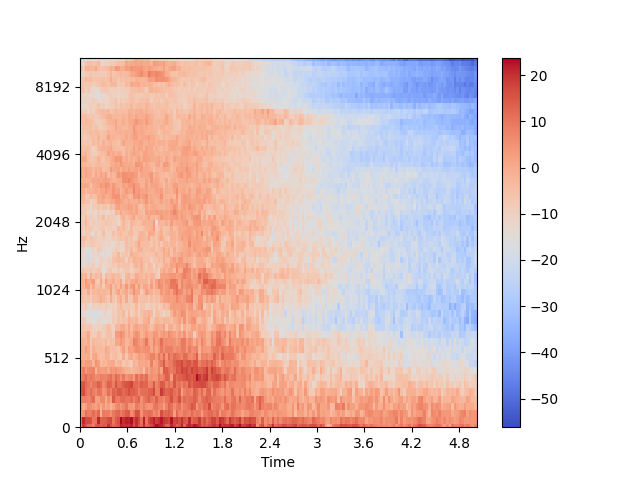}}
			\subfigure[chirping birds]{\includegraphics[width=3cm]{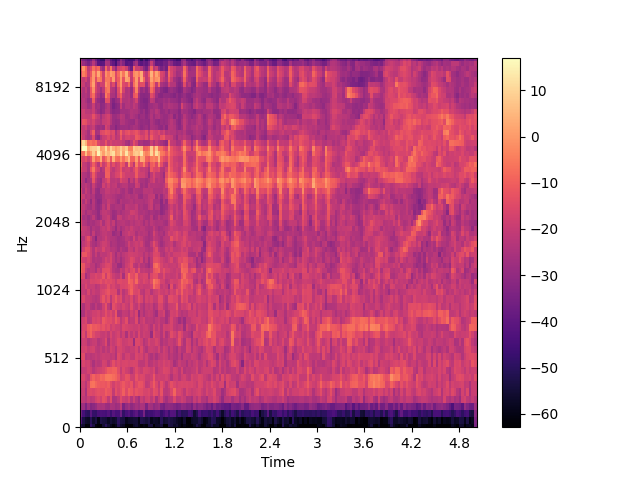}}
			\subfigure[crying baby]{\includegraphics[width=3cm]{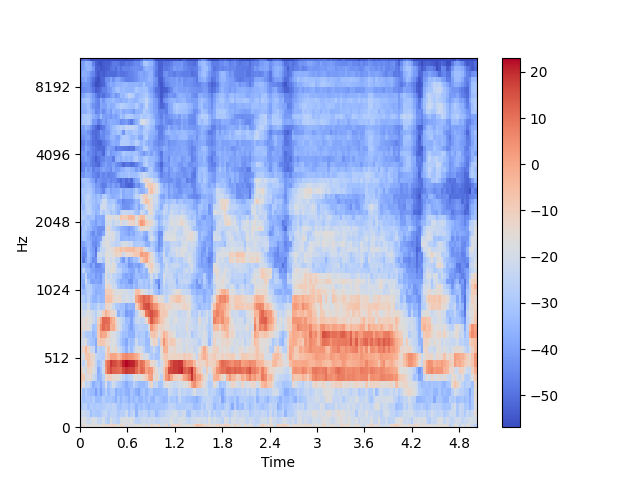}}
			\subfigure[frog]{\includegraphics[width=3cm]{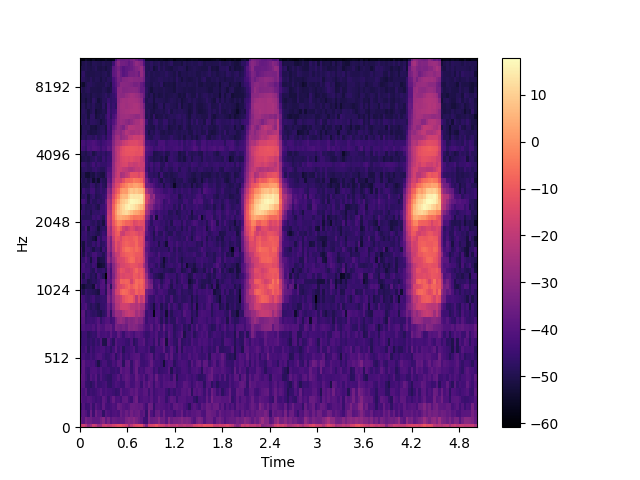}}
			\subfigure[vacuum cleaner]{\includegraphics[width=3cm]{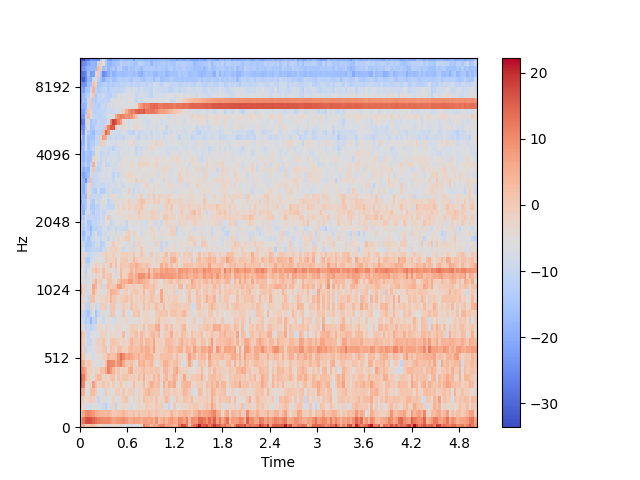}}
			\subfigure[sheep]{\includegraphics[width=3cm]{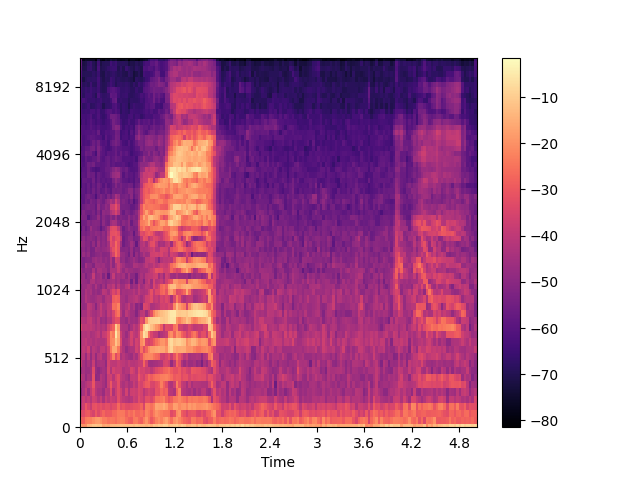}}
		\end{center}\vspace*{-10pt}\caption{Examples of log-Mel spectrograms of various environmental sounds in the ESC-50 dataset.}
		\label{fig:esc50_examples}
	\end{figure*}
	
	The classification of environmental sound is one of the most fundamental and challenging areas of audio signal processing. The ESC task is challenging because of the following two reasons. First, since environmental sounds can be generated by various machines and nature, they are very diverse, opposite to music and speech which have a common structure. Therefore, it is challenging to design an algorithm that captures the intrinsic texture and characteristics of the sound due to the dynamic and unstructured nature of the sound. Second, the acoustic signals can be viewed as the unstructured spatial layouts of various signal patches with various strengths distributed at different time stamps. Unlike the music and speech recognition tasks that contain rich semantic information, ESC has relatively limited prior information with respect to the spatial-temporal characteristics of sound signals. This makes ESC more challenging since sound signals consist of unstructured spatial-temporal correlation without explicit patterns or characteristics.
	
	To conduct the ESC task, the handcrafted features are first extracted from the sound waveform. For example, the features like the local binary pattern (LBP) \cite{Toffa2020}, Log-Mel \cite{XinyuWu2009,Wang2020}, wavelet \cite{valero2012gammatone,meintjes2018fundamental}, and Mel-frequency cepstral coefficients (MFCCs) \cite{2009Surveillance} are used. Meanwhile, there is a combination of features, e.g., the Log-Mel and MFCCs features are fused in \cite{2019Environment}. These features have been widely used and validated to be effective for ESC. In this article, following the settings in the \cite{Wang2020}, we choose the Log-Mel for preprocessing of sound signals. Then, the preprocessed features are trained with models such as support vector machine (SVM) \cite{tripathi2015acoustic} and Gaussian mixture models (GMM) \cite{dhanalakshmi2011classification} to classify environmental sounds. 
	
	Recently, benefiting from the remarkable progress of deep learning, the usage of convolutional neural networks (CNNs) to classify the spectrograms of sounds is well-exploited \cite{piczak2015environmental,Demir2020A}. CNN is powerful in extracting the discriminative feature representations on the various datasets by learning from the big training datasets.
	However, these CNN-based methods do not explicitly focus on the essential regions that lie at different locations of spectrograms of the sound waveform. Koutini et al. \cite{koutini2019receptive} propose to adjust the receptive field within a certain range of various CNN architectures which can further significantly boost the performance of ESC. Further, to capture the salient features that are more relevant to the sound events, attention mechanisms have been proposed and widely used \cite{Li2019}\cite{sharma2020environment}\cite{Wang2020}\cite{tripathi2021environment}\cite{zhang2021attention}\cite{wang2021multi}\cite{sharma2020environment}. Zhang et al. propose the frame-level attention to learn temporal correlations of sound signals on the Log Gammatone Spectrogram \cite{zhang2021attention}. Tripathi et al. \cite{tripathi2021environment} design the attention based CNN to learn the intra-class inconsistency problem. Li et al. propose temporal attention to focus on learning the diverse temporal structures of environmental sounds but they neglect the spatial distribution of sound signals. To learn the robust features of the unstructured spatial-temporal sound signals, the parallel temporal-spectral attention is designed, which considers sound signals on the spatial-temporal domains \cite{Wang2020}. However, these methods did not utilize the rich multi-scale features over the hierarchical feature maps. To remedy this issue, Demir et al. \cite{demir2020new} address multi-scale features by constructing a new pyramidal concatenated CNN, which takes three scales of Log-Mel features with average pool mechanism and hierarchical CNNs. However, its computational cost is high and it hardly addresses the spatial-temporal signals without the attention.
	
\begin{figure*}[t]
		\begin{center}
			\includegraphics[width=13cm]{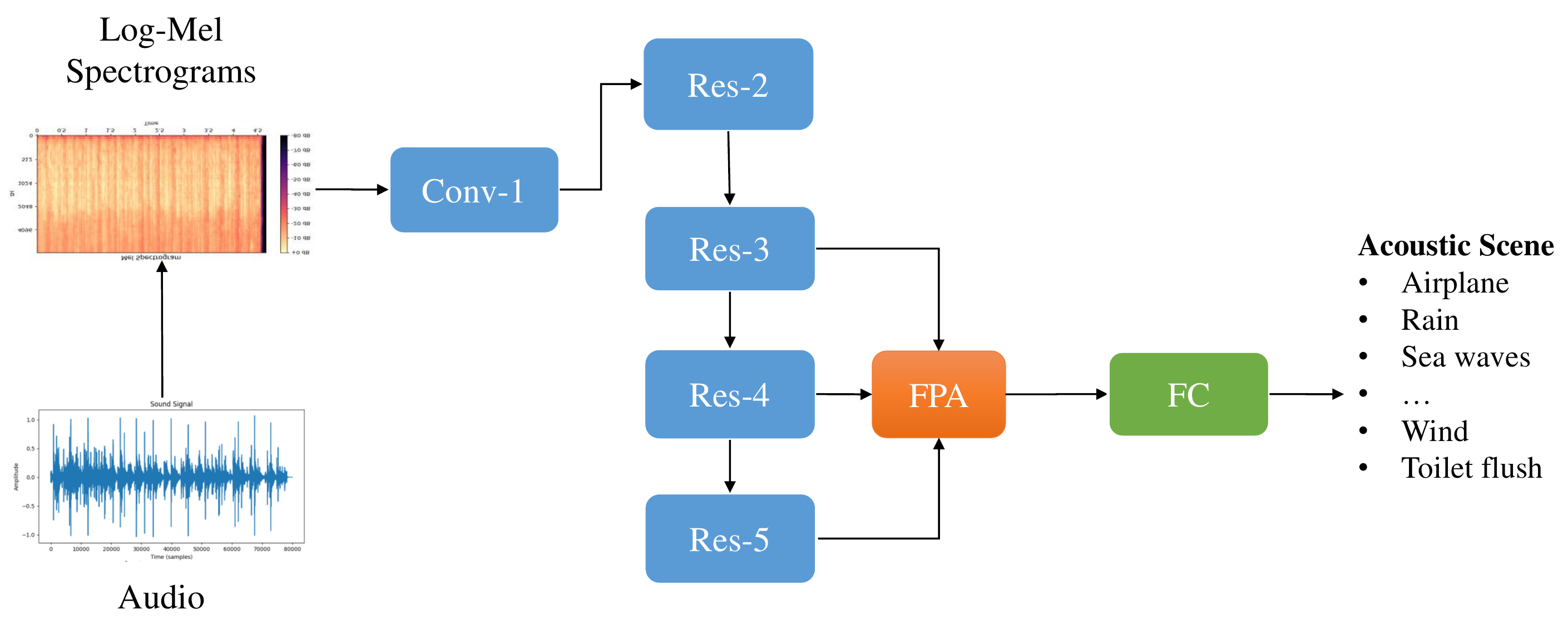}
		\end{center}\vspace*{-10pt}\caption{The proposed FPA module processes and refines the acoustic signal with a feature pyramid attention module for acoustic signal representation.}
		\label{fig:method}
	\end{figure*}

	Figure~\ref{fig:esc50_examples} shows the examples of Log-Mel spectrograms in the ESC-50 dataset. For instance, the salient and semantically relevant features of glass breaking and can opening are distributed within a short period of time, whereas the noise or silent features are distributed with a larger portion. In contrast, the salient and semantically relevant features of crying baby and chirping birds are distributed evenly across the entire period. These unstructured features and different distributions of semantically relevant features increase the difficulties in learning representations of sound spectrograms and reduce the robustness and generalization of the model. To handle such challenges, as shown in Figure~\ref{fig:method}, we propose a novel pyramid feature attention module that guides the network to focus on the more semantically relevant regions for enhanced ESC. Specially, to model the variance of sound spectrograms with a hierarchical multi-scale feature map, we leverage pyramid feature representation that can be a suitable but less exploited choice. First, we use ResNet as a backbone network that extracts features of preprocessed spectrograms. Then, to model the spectrograms of sounds in a hierarchical spatial-temporal aware manner, we construct a novel pyramid feature representation with multi-scale feature maps by aggregating the feature maps of ResNet at multiple layers with a feature pyramid attention module (FPAM). Since the multi-scale features are with different scales, we design a dimension alignment module to align the multi-scales features. Then, the processed features can be learned by SAM network individually, where the spatial information can be learned during training. To capture the temporal information, these latent features are further processed by a pyramid channel attention module. In this way, the proposed FPAM is able to learn the unstructured spatial-temporal relations of sound spectrograms. Finally, the classification results are estimated through the fully connected layer. We conduct ESC task on the ESC-50 and ESC-10 datasets to show the efficacy of the FPAM in terms of accuracy. Besides, we visualized the attention maps of the proposed FPAM, which suggest that the FPAM has the capability to learn semantically relevant regions for ESC while neglecting the silent regions.

	To summarize, our main contribution in this work can be summarized as four folds:
	\begin{itemize}
		\item We propose a novel feature pyramid attention module that refines and aggregates the features at multi-scale layers and obtains hierarchical feature representation of sound spectrogram.
		
		\item The FPAM contains SPA and CPA that refine feature representations across multiple scales, spatial, and temporal domains, thus increasing the generalization of network.
		
		\item Visualization results show the FPAM helps the network to learn the semantically relevant and salient frames and neglects the noise and silent frames.
		
		\item Experimental results obtained by the proposed method on the ESC-50 and ESC-10 are compatible with state-of-the-art methods.
		
	\end{itemize}

	The rest of the paper is organized as follows. The related works are presented in Section~\ref{sec:related_works}. The proposed FPAM network is described in Section~\ref{sec:method}. The experimental results are shown in Section~\ref{sec:exp}. The conclusion and future work about our research come in Section~\ref{sec:conclusion}.

	\section{Related Works}
	\label{sec:related_works}
	The environmental sound classification consists of two essential steps including signal preprocessing and classification. For signal preprocessing, there are various methods to preprocess the original waveform of sounds, such as Log-Mel spectrogram \cite{Wang2020}, Gammatone-like spectrogram \cite{ren2016sound}, local binary pattern \cite{Toffa2020}, wavelet transform \cite{meintjes2018fundamental}, and Mel-frequency cepstral coefficients \cite{2009Surveillance}.
	In this article, the Log-Mel spectrograms are extracted in the signal preprocessing stage. After signal preprocessing, the classifier that can well distinguish the characteristics of preprocessed acoustic signals is performed on the extracted features. With the rapid development of deep learning in image classification, numerous works have exploited the usage of convolutional neural networks (CNNs) to classify the spectrograms of sounds \cite{piczak2015environmental,zhang2018ld,demir2020new}. CNN has shown the capabilities of learning discriminative feature representations of sound signals on various datasets. However, there are still some hard samples that merely CNNs models are incapable to handle. To build feature representations that are discriminative enough for ESC, there emerged a bunch of works. Sharma et al. \cite{sharma2020environment} propose using multiple feature channels, including Mel-Frequency Cepstral Coefficients (MFCC), Gammatone Frequency Cepstral Coefficients (GFCC), the Constant Q-transform (CQT), and Chromagram for ESC with attention. Zhu et al. \cite{zhu2018learning} argue that the features extracted by single size convolutional layers are inferior for building highly discriminative representation. Hence, they build a multi-scale convolutional layer for multi-scale feature representations of sound signals. Tripathi et al. use knowledge distillation with teacher and student networks to augment learning. Guzhov1 et al. \cite{guzhov2021esresnet} propose to use feature representations from multiple domains, in which the network involves a lot of computations. 
	Recently, attention has been a powerful neural network enhancement module in deep learning based tasks and attention based methods has shown impressive performance on various tasks, including image classification \cite{woo2018cbam}, hand gesture recognition \cite{zhou2021long}, object detection \cite{zhu2018attention}, and semantic segmentation \cite{song2021attanet}. To capture the salient features that are more relevant to the sound events, attention mechanisms have been introduced \cite{Li2019} \cite{Wang2020}. For instance, temporal attention is proposed to capture the essential temporal information of sound \cite{Li2019}. A sparse key-point encoding and efficient multispike learning framework are proposed for this problem \cite{yu2020robust}. To consider the spectral information for ESC, parallel temporal-spectral attention is proposed to enhance the representation of both the temporal and spectral features by capturing the importance of different time frames and frequency bands of audios \cite{Wang2020}. 
	
	In summary, these attention mechanisms enhance the feature representations at the temporal and spectral perspectives but lack preserving the multi-scale feature representation. To this end, we propose a feature pyramid attention module to represent sound signals from multi-scale feature maps for ESC.

	\begin{figure*}[t]
		\begin{center}
			\subfigure{\includegraphics[width=13cm]{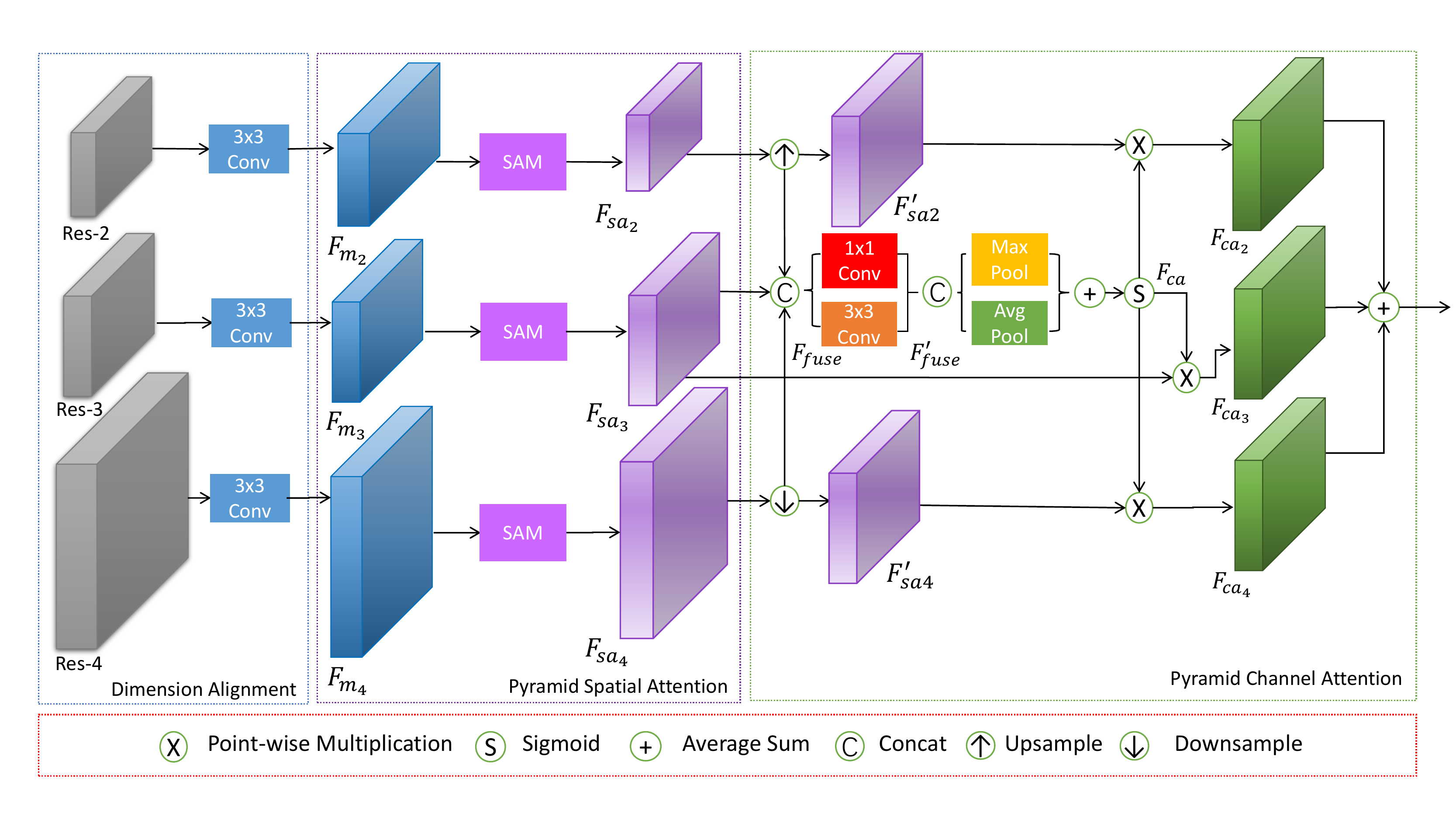}}
		\end{center}\vspace*{-10pt}\caption{Feature pyramid attention module, where the feature is first aligned in channel dimension with a dimension alignment module. Then, the feature is refined with spatial attention. Afterward, the refined feature is improved with pyramid channel attention module.}
		\label{fig:fpam}
	\end{figure*}
	
	\begin{figure}[t]
		\begin{center}
			\subfigure{\includegraphics[width=8cm]{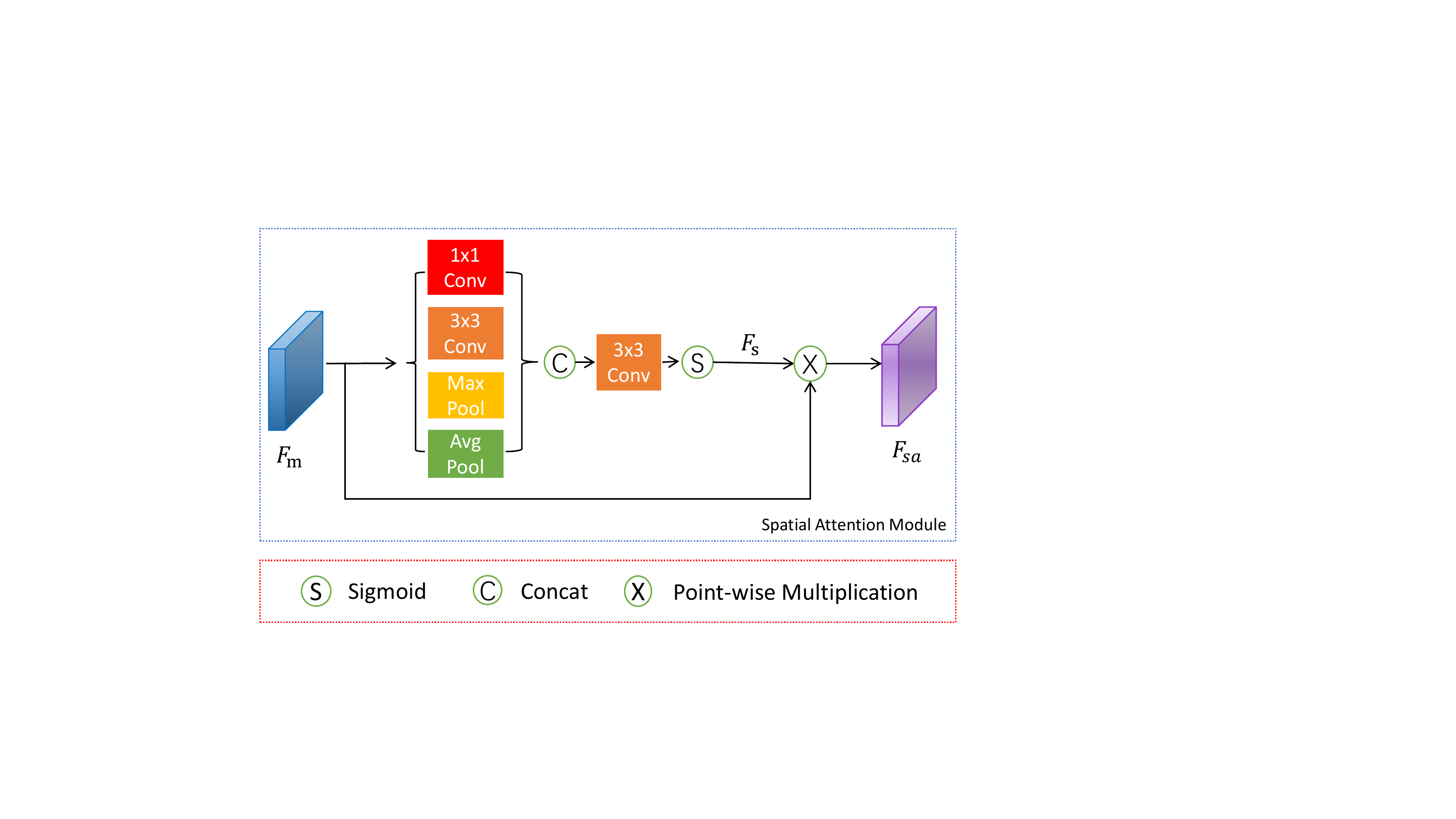}}
		\end{center}\vspace*{-10pt}\caption{Spatial attention module (SAM).}
		\label{fig:sam}
	\end{figure}

	\section{Methodology}
	\label{sec:method}
	In this section, we describe the methodology used to perform the feature pyramid attention over the multi-scale feature maps on residual neural networks. As shown in Figure~\ref{fig:method}, our proposed FPAM based residual neural network mainly contains two components. The first component is the backbone network for feature extraction, where we use the ResNet50 as the backbone network, and there are feature maps of multiple scales including Res-2, Res-3, Res-4, and Res-5. As we can see that FPAM is attached to combines the feature maps of multiple scales, and the detailed structure of FPAM network is shown in Figure~\ref{fig:fpam}. Concretely, the second component is a FPAM that refines and aggregates the pyramid-level feature maps from the backbone network. Specifically, the FPAM is composed of three modules, including dimension alignment module, pyramid spatial attention (PSA) module, and pyramid channel attention (PCA) module. The dimension alignment module is used to align the dimensions of multiple layers of backbone network. The PSA module is used for enhancement of the features along with spatial dimension, where the SAM is depicted in Figure~\ref{fig:sam}. The PCA module is used for refinement of the features along with temporal dimension.

	\subsection{Feature Pyramid Attention Module}
	The feature pyramid attention module is depicted in Figure~\ref{fig:fpam}. To exploit the semantically relevant regions of the feature maps over multiple layers during the training process, we propose a novel FPAM network to aggregate the multi-scale feature maps and to learn the essential regions of sound spectrograms. To this end, there are three modules designed, including the dimension alignment module, pyramid spatial attention module, and pyramid channel attention module. First, the feature maps from the layers of the Res-2, Res-3, and Res-4 of the backbone network ResNet50 pass through a dimension alignment module to ensure each feature map has the same channel dimension, where we obtain intermediate feature maps such as $F_{m_2}$,$F_{m_3}$, and $F_{m_4}$. Then, these feature maps are refined to obtain the semantically relevant regions spatially by the pyramid spatial attention module.
	
	The spatial attention module is depicted in Figure \ref{fig:sam}. The input feature map $F_m$ passes through a layer with 1x1 convolution, 3x3 convolution, max pooling, and average pooling. The outputs are concatenated and then refined by a 3x3 convolution. The results are sent to a Sigmoid function to obtain the spatial probabilistic feature map $F_s$, where the elements of the $F_s$ will be restricted between 0 and 1.

	\begin{equation}	
		\begin{split}
			F_{s} = \sigma ( Conv3\_3 ( Concat ( MaxPool(F_m),AvgPool(F_m), \\ Conv3\_3(F_m),Conv1\_1(F_m) ) ) ) \\
		\end{split}
	\end{equation}
	
	To obtain the spatial attention map $F_{sa}$ of the input feature, an element-wise multiplication is conducted between the input feature map $F_m$ and spatially probabilistic feature map $F_s$:
	
	\begin{equation}	
		\begin{split}
			F_{sa} = F_m \otimes F_{s}.
		\end{split}
	\end{equation}

	After that, to maintain the same dimension of output features, the output of three spatial attention modules $F_{sa2}, F_{sa3}, F_{sa4}$ will be concatenated with Upsample, the original feature map of SAM, and Downsample. Then these three features are processed by a pyramid channel attention module, where the Conv1x1, and Conv3x3 are used for the first feature refinement. The MaxPool and AvgPool are used for the second feature refinement.  
	
	\begin{equation}
		\begin{split}
			&F_{sa_2}^{'}  = Upsample(F_{sa_2})		\\		
			&F_{sa_4}^{'}  = Downsample(F_{sa_4})		\\
			&F_{fuse}  = Concat(F_{sa_2}^{'},F_{sa_3}, F_{sa_4}^{'})		\\
			&F_{fuse}^{'}  = Concat(Conv1x1(F_{fuse}), Conv3x3(F_{fuse}))				  \\
			&F_{ca}  = \sigma(MaxPool(F_{fuse}^{'})+AvgPool(F_{fuse}^{'}))				  \\
		\end{split}
	\end{equation}
	
	The output will be a channel-wise attention map $F_{ca}$. Then it is multiplied with the original feature maps to calculate the pyramid feature representation $F_{fpam}$. 
	
	\begin{equation}
		F_{fpam}  = (F_{ca}*F_{sa_2}^{'} + F_{ca}*F_{sa_3} + F_{ca}*F_{sa_4}^{'})/3
	\end{equation}

	\section{Experiments and Discussions}
	\label{sec:exp}
	In this section, we examine the performance of the proposed FPAM on the ESC task. Then, the visualization of attention maps w.r.t the sound spectrograms are analyzed, where the superior of the FPAM on the ESC task is well explained. Detailed evaluation with extensive results is shown to prove the efficacy of the proposed method.

	\begin{figure*}[htbp]
		\centering
		\subfigure[Clock Tick]{
			\begin{minipage}[t]{0.22\linewidth}
				\centering
				\includegraphics[width=3.6cm]{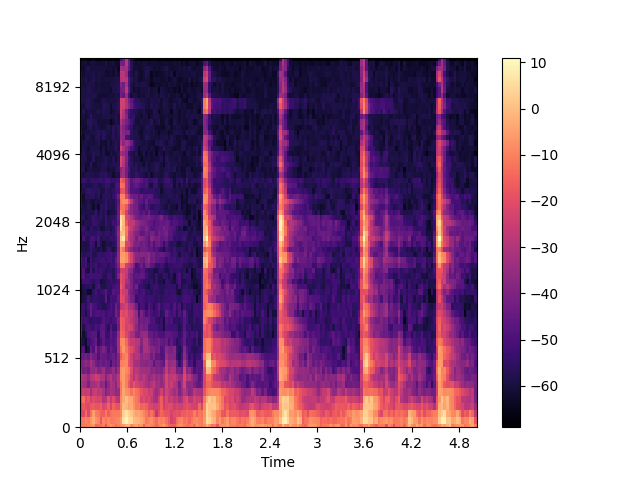}
				\vspace{0.02cm}
				\includegraphics[width=3.6cm]{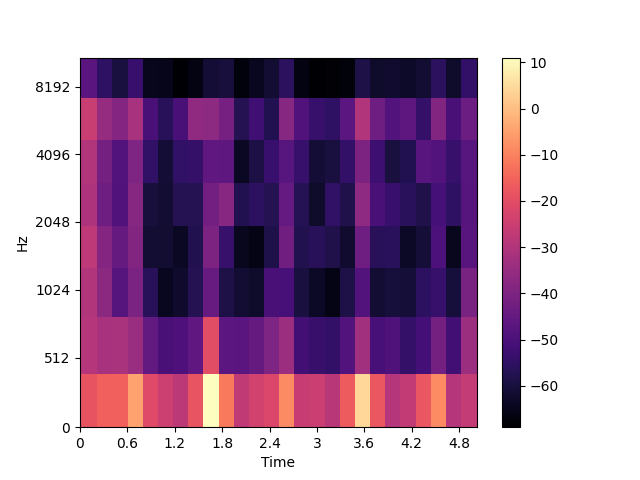} 
				\vspace{0.02cm}
			\end{minipage}%
		}
		\subfigure[Crying Baby]{
			\begin{minipage}[t]{0.22\linewidth}
				\centering
				\includegraphics[width=3.6cm]{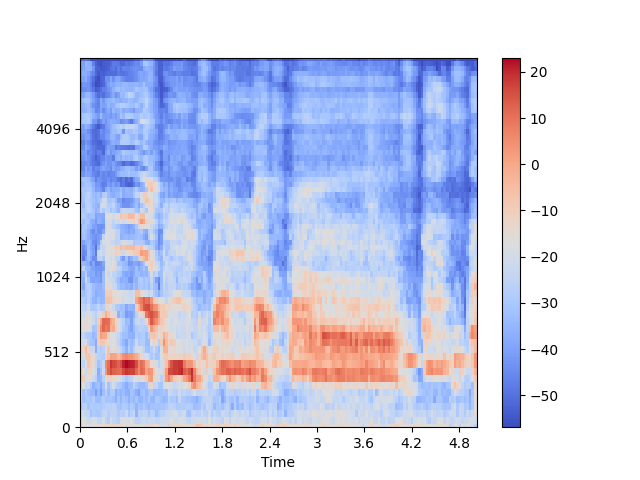}
				\vspace{0.02cm}
				\includegraphics[width=3.6cm]{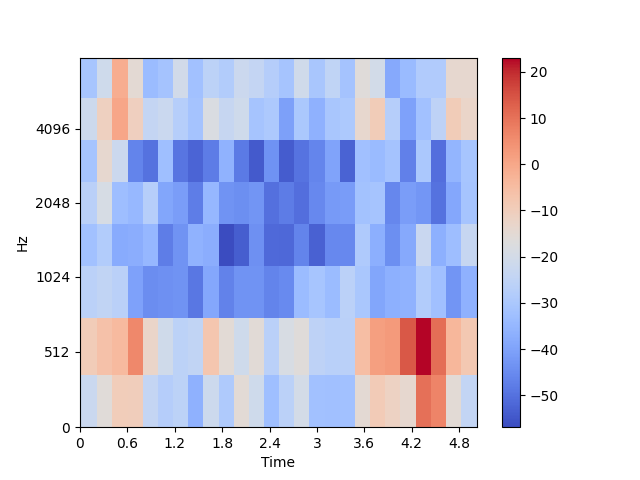}
				\vspace{0.02cm}
			\end{minipage}%
		}
		\subfigure[Dog]{
			\begin{minipage}[t]{0.22\linewidth}
				\centering
				\includegraphics[width=3.6cm]{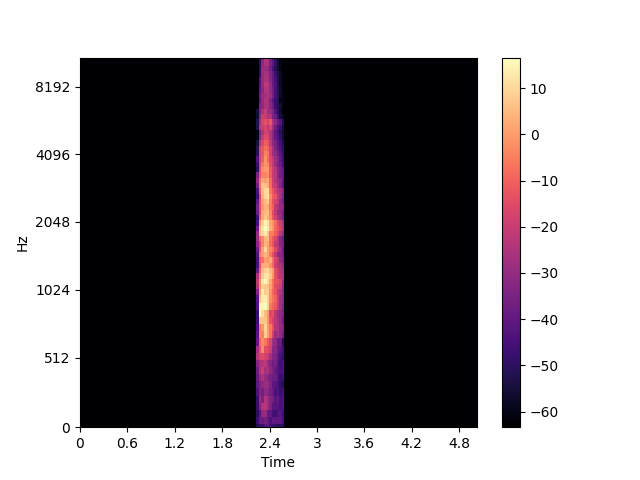}
				\vspace{0.02cm}
				\includegraphics[width=3.6cm]{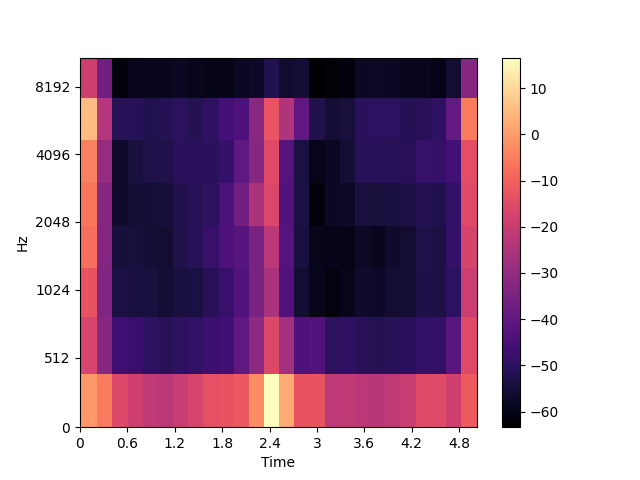}
				\vspace{0.02cm}
			\end{minipage}%
		}
		\subfigure[Sneezing]{
			\begin{minipage}[t]{0.22\linewidth}
				\centering
				\includegraphics[width=3.6cm]{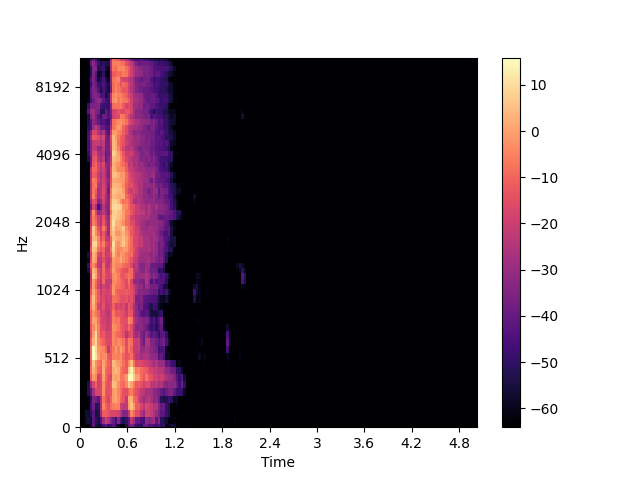}
				\vspace{0.02cm}
				\includegraphics[width=3.6cm]{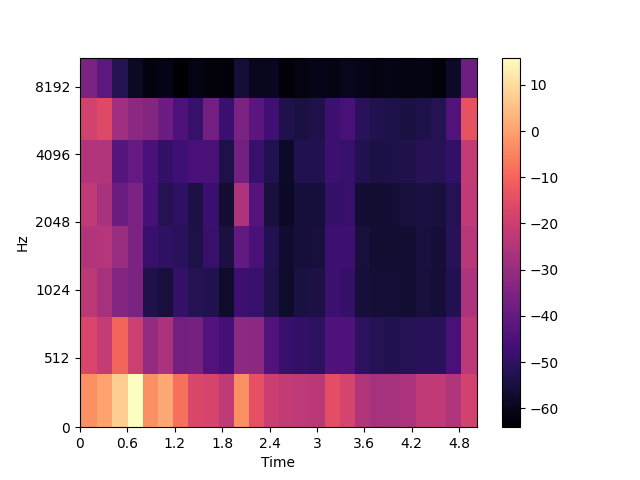}
				\vspace{0.02cm}
			\end{minipage}%
		} \\
		\subfigure[Helicopter]{
			\begin{minipage}[t]{0.22\linewidth}
				\centering
				\includegraphics[width=3.6cm]{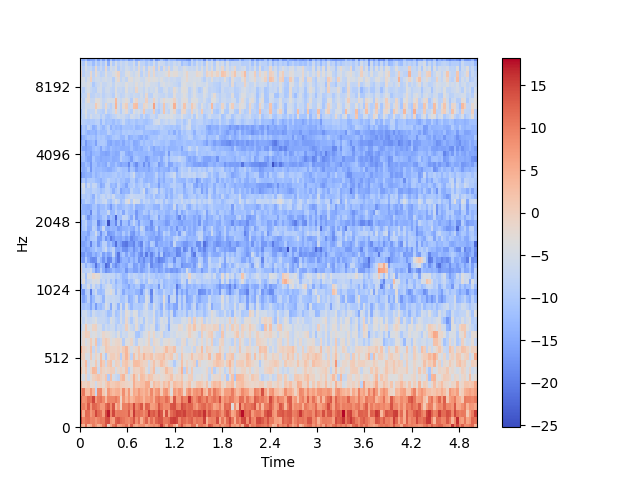}
				\vspace{0.02cm}
				\includegraphics[width=3.6cm]{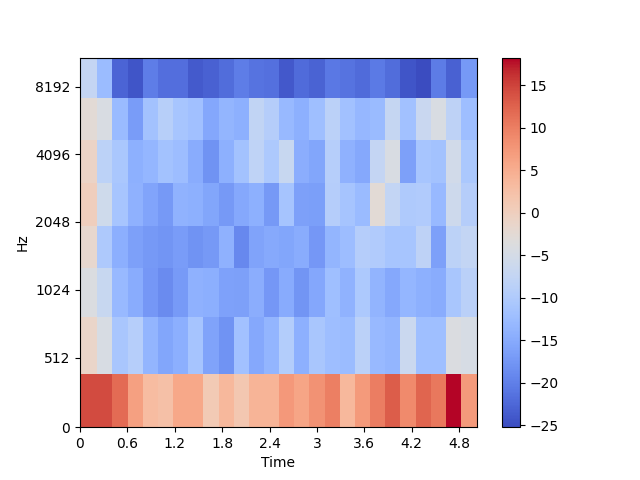}
				\vspace{0.02cm}
			\end{minipage}%
		}
		\subfigure[Church Bells]{
			\begin{minipage}[t]{0.22\linewidth}
				\centering
				\includegraphics[width=3.6cm]{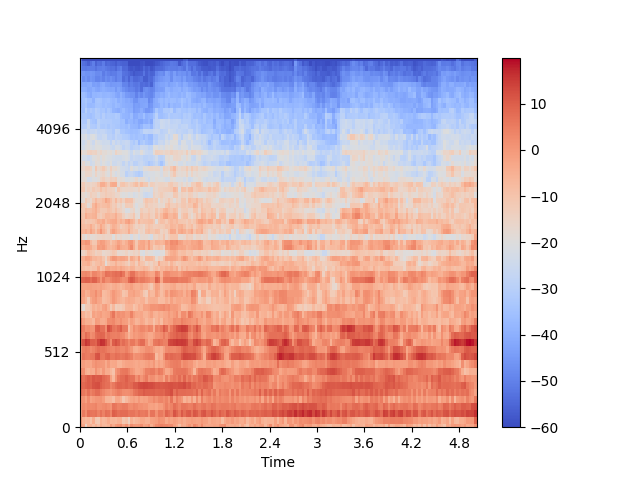}
				\vspace{0.02cm}
				\includegraphics[width=3.6cm]{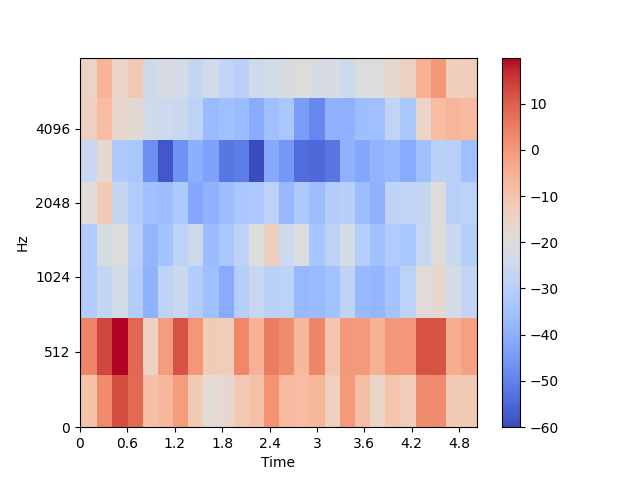}
				\vspace{0.02cm}
			\end{minipage}%
		} 
		\subfigure[Rooster]{
			\begin{minipage}[t]{0.22\linewidth}
				\centering
				\includegraphics[width=3.6cm]{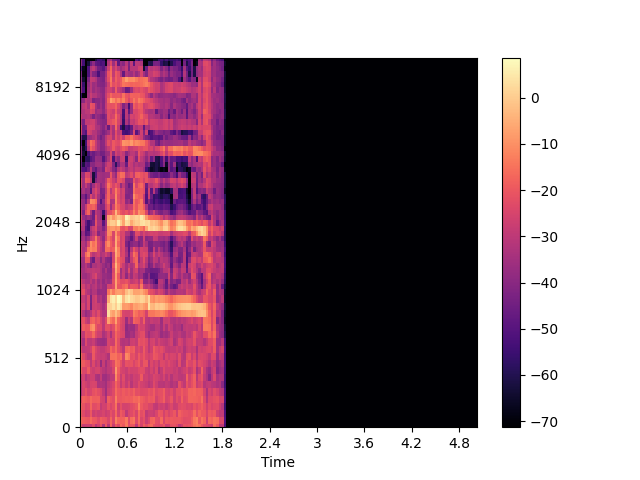}
				\vspace{0.02cm}
				\includegraphics[width=3.6cm]{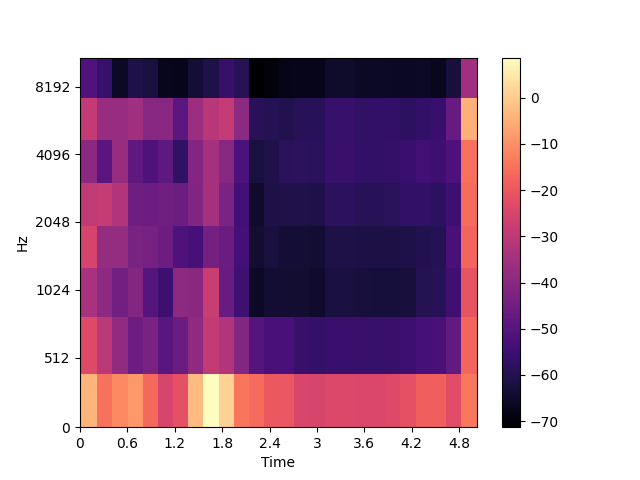}
				\vspace{0.02cm}
			\end{minipage}%
		}
		\subfigure[Rain]{
			\begin{minipage}[t]{0.22\linewidth}
				\centering
				\includegraphics[width=3.6cm]{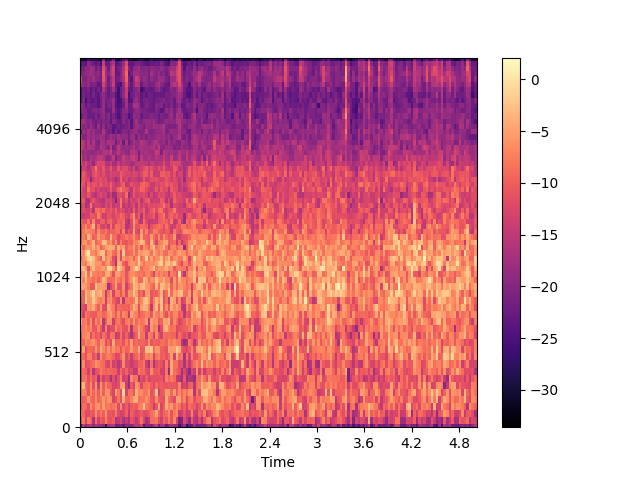}
				\vspace{0.02cm}
				\includegraphics[width=3.6cm]{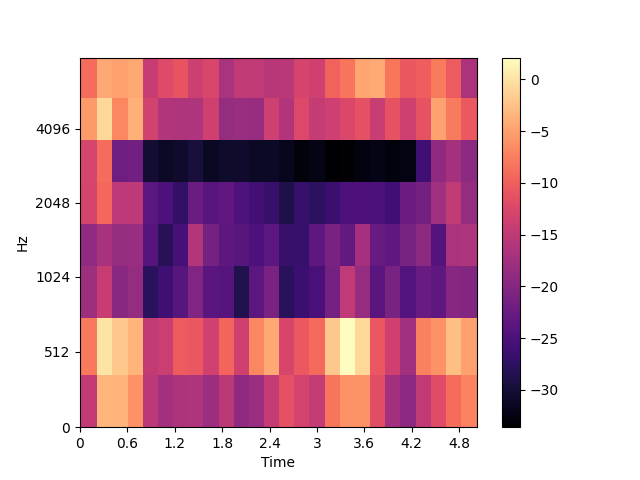}
				\vspace{0.02cm}
			\end{minipage}%
		} \\
		\centering 
		\caption{The visualization of attention maps of different sound spectrograms. The first rows are clock tick, crying baby, dog, and sneezing.	The third row are helicopter, church bells, rooster, and rain. The second and fourth row are their corresponding attention maps.}
		\vspace{-0.2cm}
		\label{fig:attention_vis}
	\end{figure*}

	\subsubsection{Datasets}
	We use Environmental Sound Classification (ESC) dataset that contains 50 acoustic scene classes for evaluation \cite{piczak2015esc}. The official dataset split settings of ESC-10 and ESC-50 are utilized, where the 5-fold cross-validation is applied. The ESC-50 dataset contains 2000 labeled environmental sounds with a duration of 5 seconds, with each category being equally distributed. The ESC-50 has a training set with 1600 samples and a test set with 400 samples. The ESC-10 is a subsect of ESC-50 that contains 320 samples as a training set, and 80 samples as a test set.
	
	\subsubsection{Implementation Details}
	The raw audios are resampled to 16.0kHz and fixed to a certain length, 5s for ESC-10 and ESC-50. Then, to analyze the signal in the frequency domain, the short-time Fourier transform (STFT) is applied to the audio signals, where the spectrograms are obtained with a hop length of 400 and window length of 1024. Last, to obtain the Log-Mel spectrograms, the 64 mel filter banks are applied to the previous spectrograms and followed by a logarithmic operation. During the training stage, we adopt the ResNet50 pretrained on the AudioSet for sound information extraction \cite{wang2018polyphonic}. The proposed model is implemented in Pytorch \cite{paszke2019pytorch}. The SGD optimizer is used for training in experiments. The initial learning rate is 0.01 and decreased by a factor of 10 at every 20 epochs. The momentum of the optimizer is 0.9. The total train epochs are 60. The network structure in detail is listed in Table~\ref{tab:FPAM-ESC}. Since the input feature of the sound spectrogram after Fourier transform is of shape 1x201x64, we adjust the input channel of ResNet50 from 3 to 1.

	\begin{table}[]
		\centering
		\caption{Network details of FPAM for environmental sound classification scene recognition based on ResNet50 baseline model.  Omit the batch size.}
		\label{tab:FPAM-ESC}
		\begin{tabular}{c|c|c}
			\hline
			Layer   & Input/Input Size & Output Size \\ \hline
			Conv1   &  1x201x64    &  64x101x32         \\ \hline
			Res-2   &  64x101x32   &  256x101x32           \\ \hline
			Res-3   &  256x101x32  & 512x51x16 \\ \hline
			Res-4   &  512x51x16   & 1024x26x8  \\ \hline
			Res-5   &  1024x26x8   & 2048x13x4  \\ \hline
			FC      &    2048x13x4 & 2048  \\ \hline
			FPAM    &  Res-3, Res-4, Res-5   & 1024x26x8  \\ \hline
			Mean    &  1024x26x8   & 1024x1  			  \\ \hline
			FC 		& 1024 & \# Class\ \\ \hline
		\end{tabular}
	\end{table}

	\subsubsection{Comparison with stat-of-the-arts}
	Our method is evaluated on two ESC datasets, ESC-10 and ESC-50, which are the most commonly used datasets for ESC, where the Log-Mel spectrograms are extracted from the audio signals as the input of the FPAM. We compare our proposed model with existing state-of-the-art methods. Table~\ref{tab:esc} demonstrates the performance of the proposed FPAM on the ESC-10 and ESC-50 datasets. The proposed FPAM reaches a comparable performance over the state-of-the-art methods on both datasets. In the ESC-10 dataset, our method reaches 99.3\% classification accuracy, which is about 3.5\% higher than TS-CNN10. In the ESC-50 dataset, our method reaches 91.6\% classification accuracy, which is 3.0\% higher than TS-CNN10.
	
	\begin{table}[]
		\centering
		\caption{Comparison of the FPAM and existing state-of-the-art methods on the ESC-10, ESC-50 datasets. We perform 5-fold cross-validation followed by official fold settings. The average accuracy of 5-fold cross-validation are reported.}
		\label{tab:esc}
		\begin{tabular}{lcc}
			\hline
			Model & ESC-10 & ESC-50 \\ \hline
			KNN \cite{piczak2015esc} & 66.7 & 32.3 \\ 
			SVM \cite{piczak2015esc} & 67.5 & 39.6 \\ 
			Random Forest \cite{piczak2015esc} & 72.7 & 44.3 \\ 
			AlexNet \cite{boddapati2017classifying} & 78.4 & 78.7 \\ 
			Google Net \cite{boddapati2017classifying} & 63.2 & 67.8 \\ 
			PiczakCNN \cite{piczak2015esc} & 80.5 & 64.9 \\ 
			SoundNet \cite{aytar2016soundnet} & 93.7 & 79.1 \\ 
			Multi-Stream CNN \cite{Li2019} & 94.2 & 84.0 \\ 
			ACRNN \cite{zhang2021attention} & 93.7 & 86.1 \\ 
			MelFB+LGTFB+EN-CNN \cite{park2020cnn} & 93.7 & 88.1 \\
			Attention Network\cite{Li2019} & 94.2 & 84.0 \\
			Human\cite{piczak2015environmental} & 95.7 & 81.3 \\
			TS-CNN10 \cite{Wang2020} & 95.8 & 88.6 \\ 
			ARNN \cite{tripathi2021environment} & 92.0 & - \\
			MCTA-CNN \cite{wang2021multi} & 94.5 & 87.1 \\
			DA-KL \cite{tripathi2022data} & 92.5 & - \\
			ACRNN \cite{zhang2021attention} & 93.7 & 86.1 \\ \hline
			
			FPAM (Ours) & \textbf{99.3}  & \textbf{91.6} \\ \hline
		\end{tabular}
	\end{table}

	\subsection{Efficacy of Attention Module}
	To show the effectiveness of the proposed FPAM, we have conducted the ablation study. From Table~\ref{tab:fpam_abl}, we observed that the proposed FPAM greatly improves the performance compared with the baseline model under the same setting. The performance of FPAM on the ESC-10 dataset is 2.8\% higher than the baseline model. The performance of FPAM is 4.6\% higher than the baseline model on the ESC-50 dataset. The Mixup \cite{zhang2017mixup} data augmentation technology is also used during training in the ablation study. However, it does not enhance the performance of the proposed FPAM. Therefore, the Mixup is not included in the final model. 

\begin{figure}[t]
		\begin{center}
			\subfigure{\includegraphics[width=4cm]{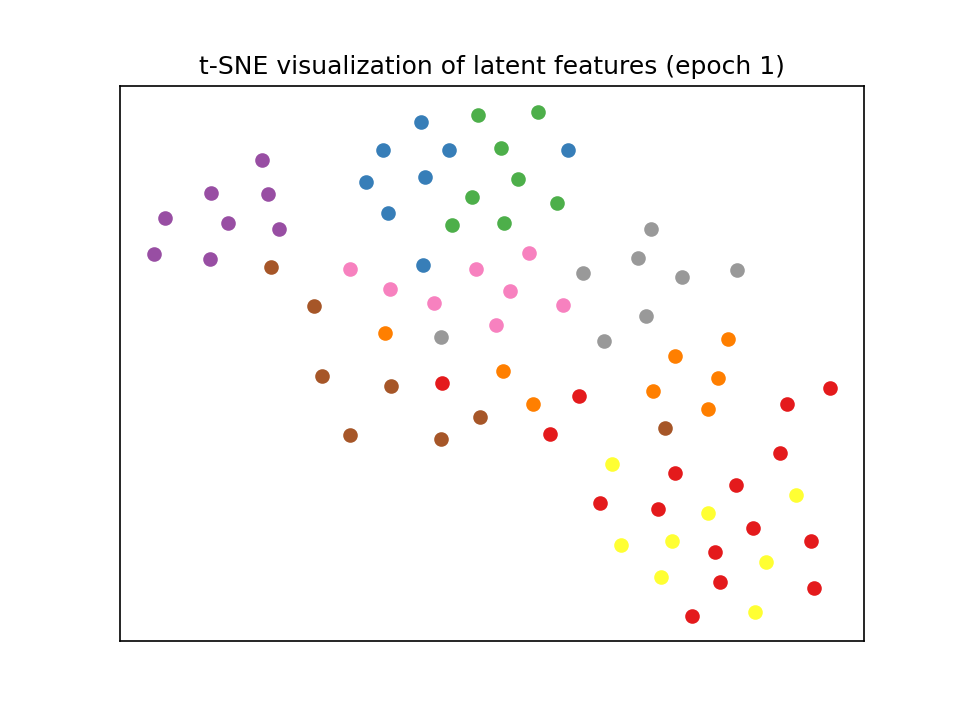}}
			\subfigure{\includegraphics[width=4cm]{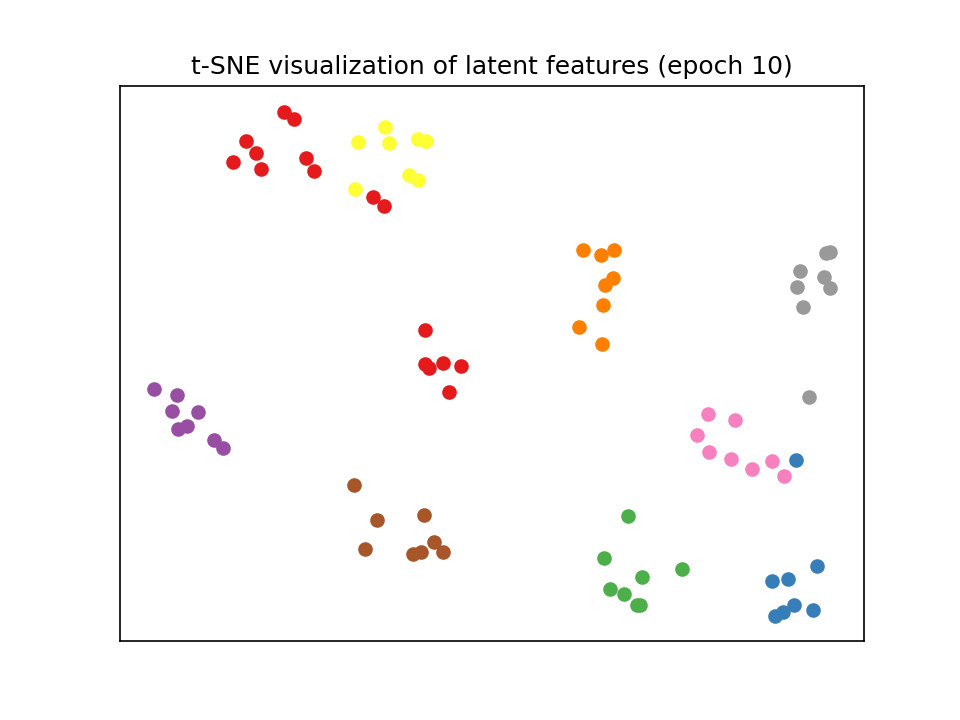}}
			\subfigure{\includegraphics[width=4cm]{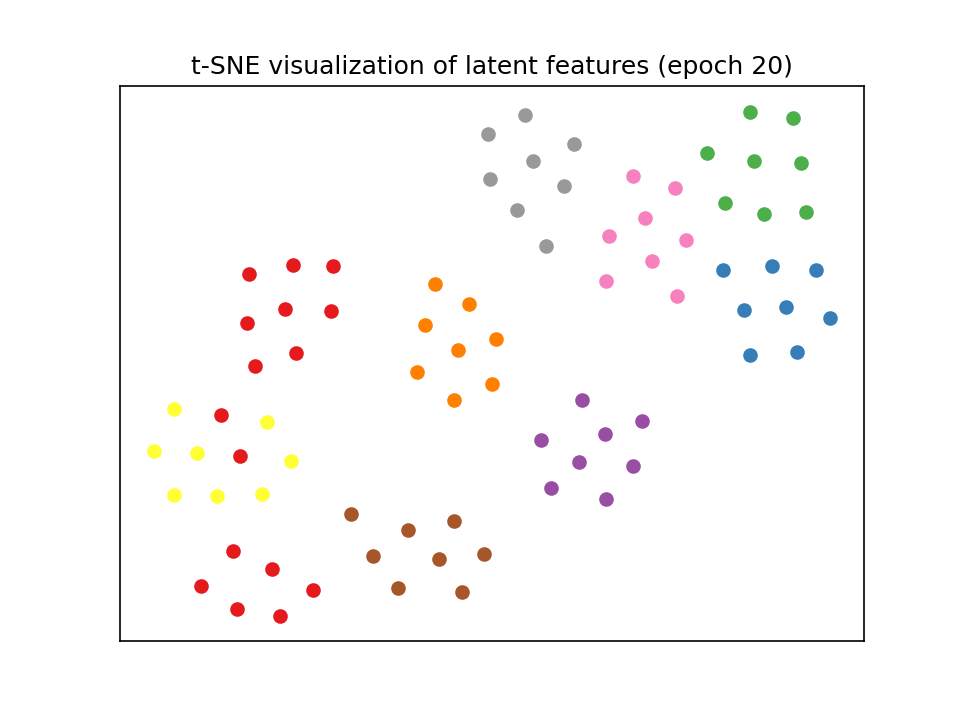}}
			\subfigure{\includegraphics[width=4cm]{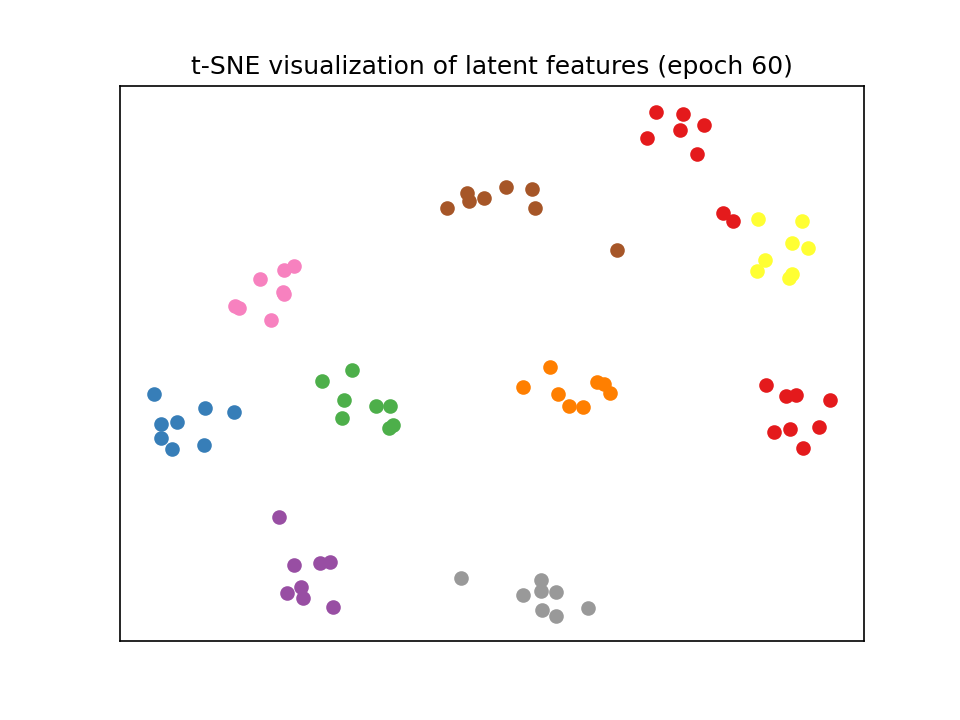}}
		\end{center}\vspace*{-10pt}\caption{t-SNE visualization of latent features after first epoch, 10-th epoch, 20-th epoch and last epoch for ESC-10 dataset.}
		\label{fig:t-SNE}
	\end{figure}

\begin{figure}[t]
		\begin{center}
			\subfigure{\includegraphics[width=4cm]{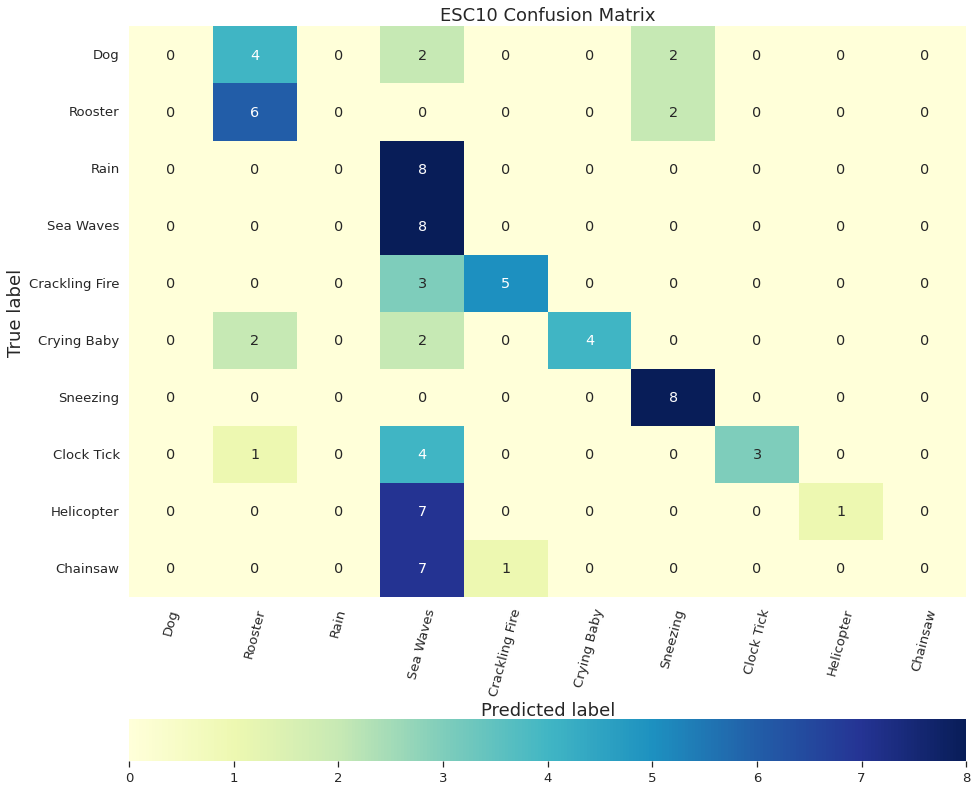}}
			\subfigure{\includegraphics[width=4cm]{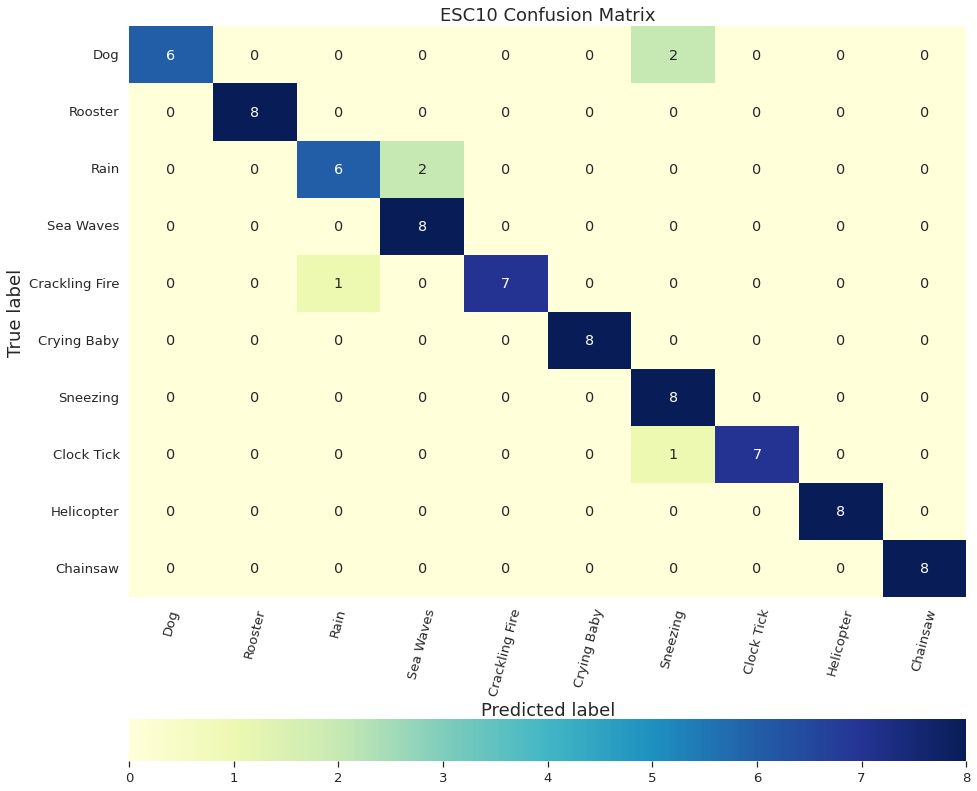}}
			\subfigure{\includegraphics[width=4cm]{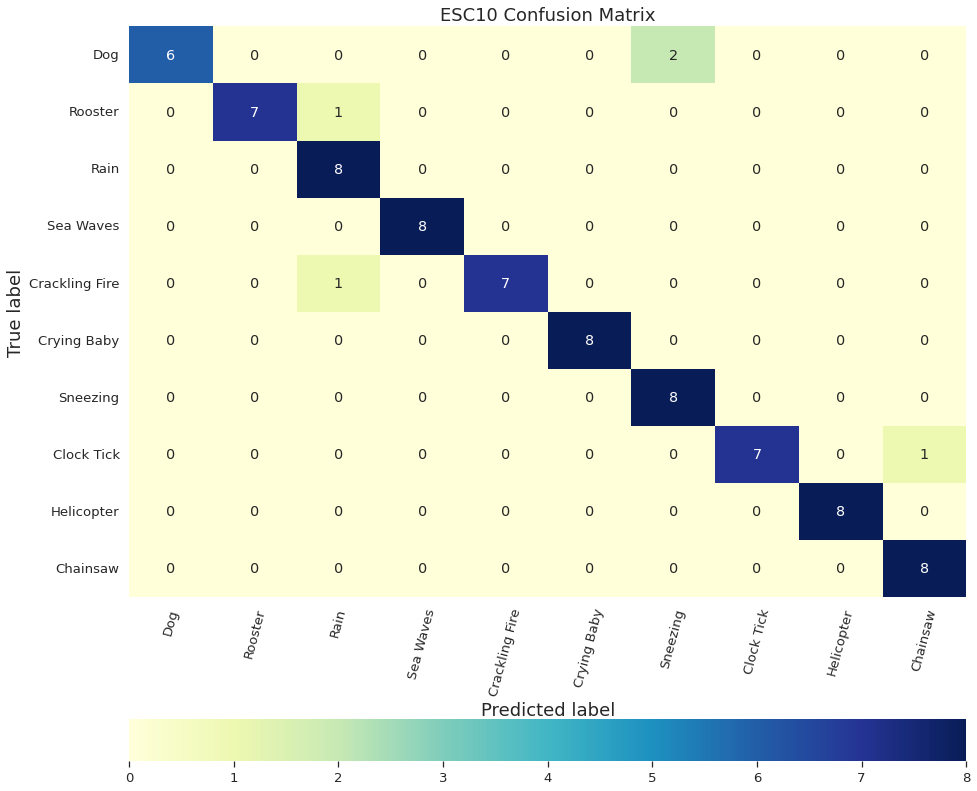}}
			\subfigure{\includegraphics[width=4cm]{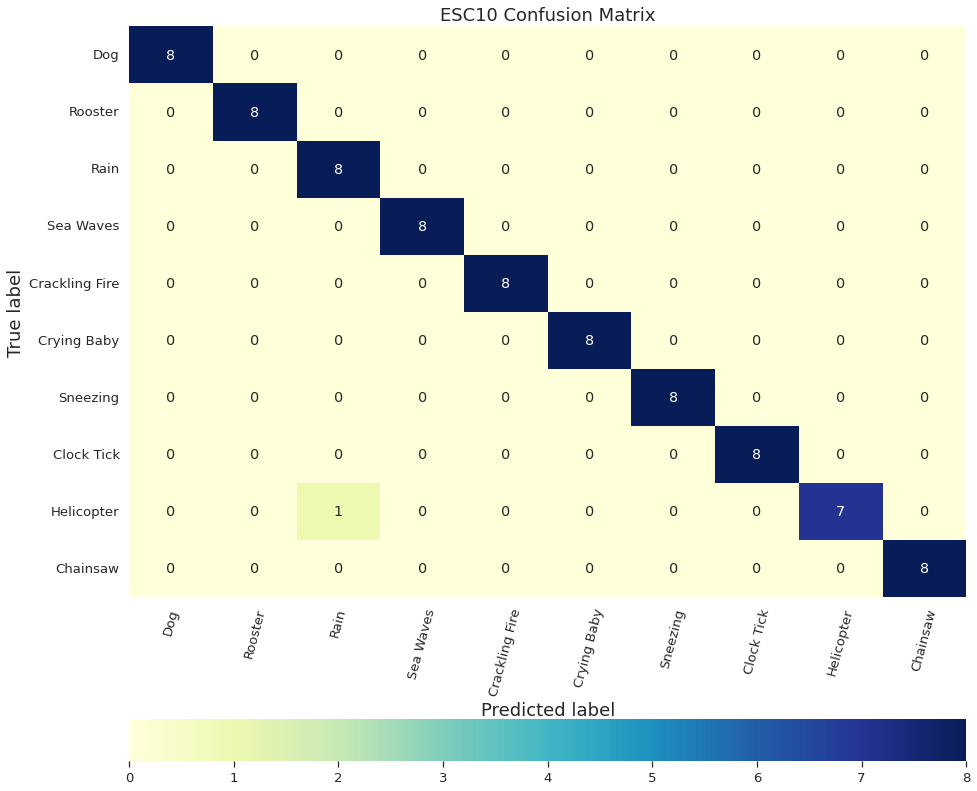}}
		\end{center}\vspace*{-10pt}\caption{Confusion Matrix of accuracy on the ESC-10 dataset. The first epoch, 5-th epoch, 10-th epoch, and last epoch are displayed.}
		\label{fig:conf_mat_esc10}
	\end{figure}

\begin{figure*}[t]
		\begin{center}
			\subfigure{\includegraphics[width=12cm]{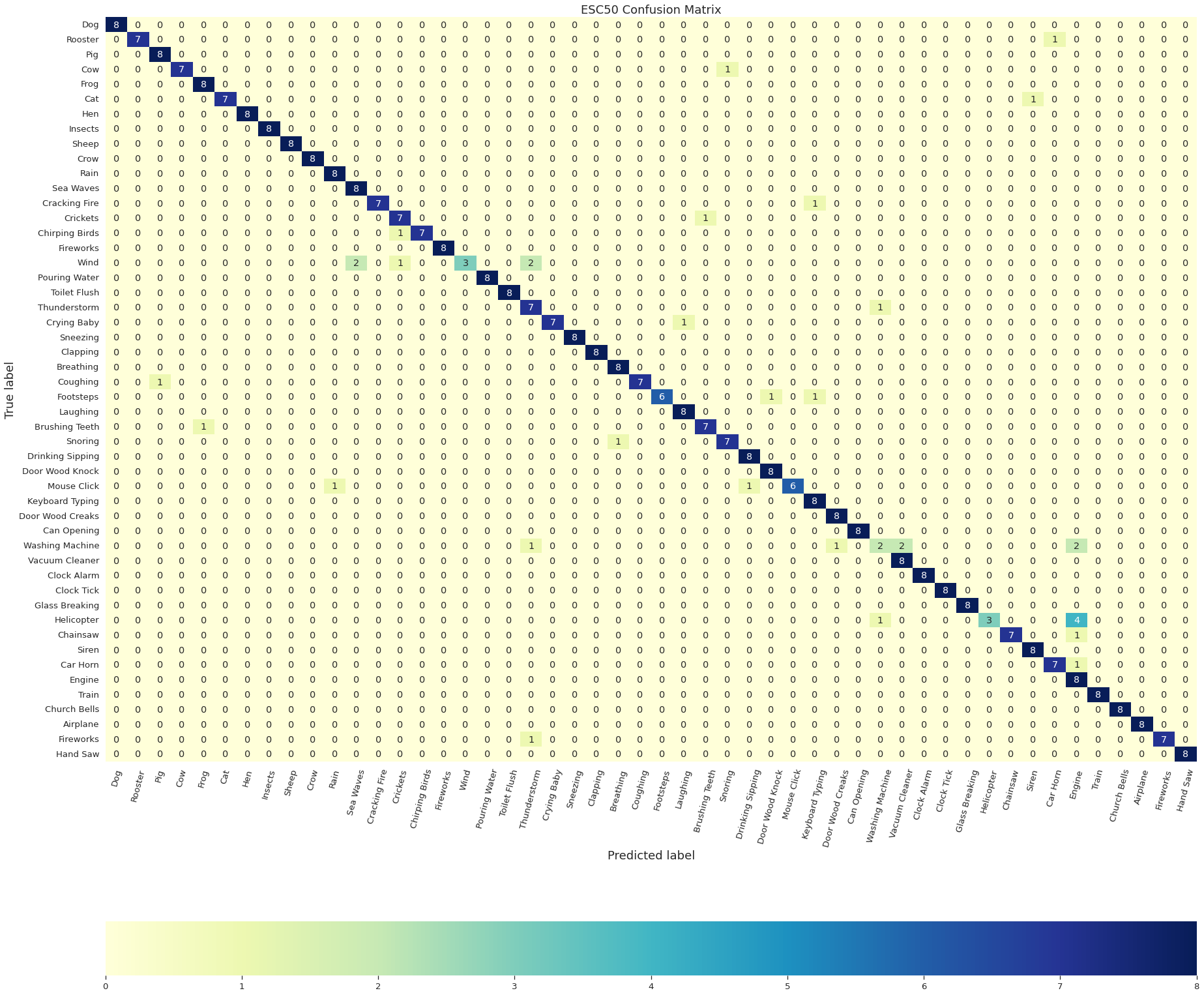}}
		\end{center}\vspace*{-10pt}\caption{Confusion Matrix for ESC-50 dataset.}
		\label{fig:conf_mat_esc50}
	\end{figure*}

	Besides, the t-SNE visualization of latent features is shown in Figure~\ref{fig:t-SNE}. At the first epoch, the latent features are distributed over the space without clear boundaries and the recognition accuracy is relatively low at this point. It is obvious that with the increase of training epochs, the latent features learned by proposed methods are distributed over the high dimensional with distinct boundaries for classification. Moreover, Figure~\ref{fig:conf_mat_esc10} shows the detailed confusion matrix of proposed method on ESC-10 dataset. As the training epoch increases, the classification accuracy is much improved. We notice that at the last epoch, there is only one error from Helicopter. These results suggest that the robust classifier is learned with the FPAM. Figure~\ref{fig:conf_mat_esc50} shows the confusion matrix of the ESC-50 dataset. We see that most classes reach accuracy higher than 87.5\%(7/8). There are three classes with a relatively lower accuracy 37.5\% (3/8). For example, there is wind (3/8) being misclassified as sea waves, thunderstorms, and so on. There is the washing machine misclassified as the vacuum cleaner, clock alarm, engine, etc. There is the helicopter misclassified as engine and washing machine. The misclassified examples are mostly within parent categories, that is, the wind, sea waves, and thunderstorms are all belong to natural weather phenomenon.
	
\begin{figure}[t]
		\begin{center}
			\subfigure{\includegraphics[width=3.5cm]{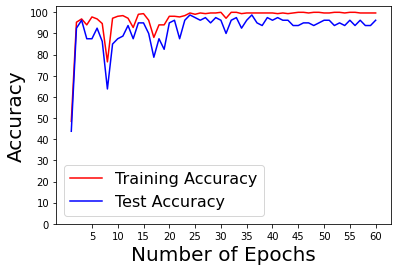}}
			\subfigure{\includegraphics[width=3.5cm]{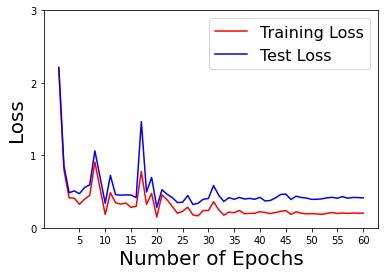}}
			\subfigure{\includegraphics[width=3.5cm]{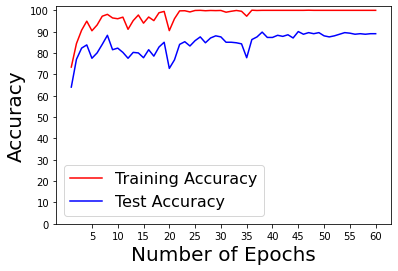}}
			\subfigure{\includegraphics[width=3.5cm]{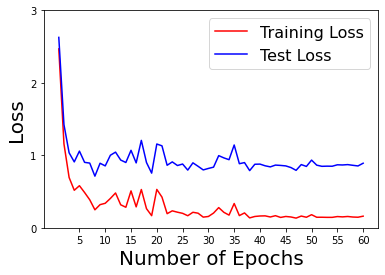}}
		\end{center}\vspace*{-10pt}\caption{Curves of accuracy and loss during the training process. The first row is the experiment on the ESC-10 dataset, while the second row is the experiment on the ESC-50 dataset.}
		\label{fig:train_val_curve}
	\end{figure}

	The curve of training and test process on both accuracy and loss shown in Figure \ref{fig:train_val_curve} also validates that the classification accuracy is increased with the increase of training epochs. During the training process, the training accuracy is always better than test accuracy, and the training loss is always smaller than the test loss. Besides, the loss is converged as the number of epochs increases, where the rate is very fast in the first ten epochs. These phenomena suggest that the network training process is smooth and healthy.

	\begin{table}[]
		\centering
		\caption{Ablation study of the proposed FPAM and its corresponding configurations with/without mix-up}
		\label{tab:fpam_abl}
		\begin{tabular}{lccc}
			\hline
			Model 	   & M/U & ESC-10 & ESC-50 \\  \hline       
			Baseline   &  \XSolidBrush &  97.0  & 86.2         \\
			FPAM  	   & \XSolidBrush  & \textbf{99.3} & \textbf{91.6} \\  \hline
			Baseline & \Checkmark  & 95.7   &  85.9 \\
			FPAM & \Checkmark   & 98.5 & 90.5		 \\		\hline
			
		\end{tabular}
	\end{table}


	\subsubsection{Visualization of Attention Maps}
	
	To further justify the efficacy of the proposed FPAM on the ESC, we show the visualization of attention maps obtained by the proposed FPAM. In Figure~\ref{fig:attention_vis}, the first/third row shows the spectrograms of sound signals while the second/fourth row shows the corresponding attention maps obtained by the FPAM.
	
	The clock tick is with repetitive patterns and is well represented by the attention map. The dog bark is with a rapid and sharp pattern, which is also well represented by the attention maps, and almost all the silent frames are filtered out. As shown in 3rd column, the sound of the rooster is a long and sharp pattern with more than half silent time frames. The visualized attention map shows the well-captured period of the rooster sound while abandoning those silent frames. It can be seen that the proposed FPAM makes use of the most discriminative signal bands while the silent part of signal has been abandoned.
	
	Therefore, the visualization analysis of attention maps shows how our method focuses on the semantically relevant regions of sound spectrograms. Besides, it shows the salient and irrelevant frames are ignored by the FPAM. These phenomena indicate that FPAM well exploited the semantically relevant textures and frequency bands and abandoned the less relevant information for spectrogram representation.
	\section{Conclusion}
	\label{sec:conclusion}
	In this paper, we propose the FPAM network for sound signal representation that exploits the salient and semantically relevant frames of sound signals. We explicitly construct the feature pyramid representations with FPAM to focus on the important regions of spectrograms of sound waveform. The extensive experiments have been conducted on two widely used datasets for environmental sound classification: the ESC-10 and ESC-50. These experiments show the efficacy of the proposed methods in two folds. First, the visualization experiment shows that the FPAM helps the network to learn the semantically relevant and salient regions of sound spectrograms. Meanwhile, we show that FPAM achieves the state-of-the-art performances on the ESC.
	
	\section{CRediT authorship contribution statement}
	\textbf{Liguang Zhou}: Conceptualization, Methodology, Software, Writing - original draft, Investigation \textbf{Yuhongze Zhou}:  Software, Visualization, Validation, Writing - review \& editing	\textbf{Xiaonan Qi}:  Software, Validation, Writing - original draft \textbf{Junjie Hu}: Validation, Writing - review \& editing 	\textbf{Tin Lun Lam}: Conceptualization, Validation, Writing -  review \& editing 	\textbf{Yangsheng Xu}: Conceptualization, Validation, Writing - review \& comment
	
	\section{Declaration of Competing Interest}
	The authors declare that they have no known competing financial interests or personal relationships that could have appeared to influence the work reported in this paper.
	
	\section*{Acknowledgment}
	This work was supported by the funding AC01202101103 from the Shenzhen Institute of Artificial Intelligence and Robotics for Society.

	
	\newpage
	\bibliography{scene_recognition}

\end{document}